\documentclass[mathleft]{an_my}
\pdfoutput=1
\usepackage{graphicx}
\usepackage{amssymb}
\usepackage{times}
\usepackage{natbib}
\bibpunct{(}{)}{;}{a}{}{}
\usepackage{journal_shortcuts}
\begin{document}

\Pagespan{1}{}
\Yearpublication{2009}%
\Yearsubmission{2009}%
\Month{99}%
\Volume{999}%
\Issue{99}%

\newcommand{\degree}{^o}
\newcommand{\K}{\,\textrm{K}}
\newcommand{\Kpc}{\,\textrm{kpc}}
\newcommand{\Mpc}{\,\textrm{Mpc}}
\newcommand{\PC}{\,\textrm{pc}}
\newcommand{\Yr}{\,\textrm{yr}}
\newcommand{\CM}{\,\textrm{cm}}
\newcommand{\Myr}{\,\textrm{Myr}}
\newcommand{\Gyr}{\,\textrm{Gyr}}
\newcommand{\Kms}{\,\textrm{km}\,\textrm{s}^{-1}}
\newcommand{\Cm}{\,\textrm{cm}}
\newcommand{\Erg}{\,\textrm{erg}}
\newcommand{\ccm}{\,\textrm{cm}^{-3}}
\newcommand{\gccm}{\,\textrm{g}\,\textrm{cm}^{-3}}
\newcommand{\Presunit}{\,\textrm{erg}\,\textrm{cm}^{-3}}
\newcommand{\MicroG}{\,\mu\textrm{G}}

\newcommand{\Sun}{_{\sun}}
\newcommand{\Rot}{_{\mathrm{rot}}}
\newcommand{\Max}{_{\mathrm{max}}}
\newcommand{\Min}{_{\mathrm{min}}}
\newcommand{\Gal}{_{\mathrm{gal}}}
\newcommand{\ICM}{_{\mathrm{ICM}}}
\newcommand{\DM}{_{\mathrm{DM}}}
\newcommand{\Gas}{_{\mathrm{gas}}}
\newcommand{\Stars}{_*}
\newcommand{\Bulge}{_{\mathrm{bulge}}}
\newcommand{\Ram}{_{\mathrm{ram}}}
\newcommand{\KH}{_{\mathrm{KH}}}
\newcommand{\Grav}{_{\mathrm{grav}}}

\title{Ram pressure stripping of disk galaxies in galaxy clusters}

\author{E. Roediger
\thanks{Corresponding author:
  \email{e.roediger@jacobs-university.de}\newline}
}
\titlerunning{Ram pressure stripping in galaxy clusters}
\authorrunning{E. Roediger}
\institute{
Jacobs University Bremen, PO Box 750\,561, 28725 Bremen
}

\received{11 August 2009}
\accepted{11 Nov 2005}
\publonline{later}


\abstract{%
While galaxies move through the intracluster medium of their host cluster, they experience a ram pressure which removes at least a significant part of their interstellar medium. This ram pressure stripping appears to be especially important for spiral galaxies: this scenario is a good candidate to explain the differences observed between cluster spirals in the nearby universe and their field counterparts. Thus, ram pressure stripping of disk galaxies in clusters has been studied intensively during the last decade. I review advances made in this area, concentrating on theoretical work, but continuously comparing to observations.}

\maketitle

\section{Introduction}
Galaxies populate different environments, ranging from the field, characterised by a low galaxy number volume density, to densely populated clusters. In nearby clusters, especially cluster spiral galaxies differ from their counterparts in the field in a number of properties (see also the extensive review by \citealt{Boselli06review} and references therein): 
\begin{itemize}
\item cluster galaxies are HI deficient compared to their field counterparts. The deficiency increases towards the cluster centre. Spatially resolved studies reveal that the HI deficiency is caused by a truncation of the gas disks. While the HI disks  of field spirals typically extend beyond the optical disks, the opposite is true for HI deficient spirals. 
\item  Luminous cluster spirals have, on average, a lower star formation rate (SFR). The suppression of star formation (SF) goes hand in hand with HI deficiency. Spatially resolved studies reveal that e.g. the H$\alpha$ disks are also truncated. 
\item Cluster spirals tend to be redder than field spirals, also indicating less active SF. 
\item An increased 20 cm radio continuum intensity suggests an increase in magnetic field  (MF) strength by factor of 2-3 compared to field spirals. 
 \item late-type galaxies follow more radial orbits and  tend to have higher velocites than early-type galaxies, which suggests they are free-falling into cluster. 
\end{itemize}
The global contents of molecular gas seems to be comparable between field and cluster spirals, the investigation of the global dust contents is difficult and still controversal.

All taken together, these observations suggests that one or more processes in cluster environments remove gas from galaxies or make them consume their gas, which leads to a  subsequent decrease of SF activity and hence change in colour. A number of processes are suspected to be responsible: \citet{Larson80} suggested the starvation scenario: With a moderate SFR, a typical spiral galaxy consumes its disk gas within a few Gyr. Dense environments cut the disk galaxies off its external gas supply, e.g. their halos, and thus SF ceases.
Tidal interactions or mergers are more common in galaxy group environments, and have a profound impact on the affected galaxies. In clusters, encounters between galaxies happen at a much higher velocity, thus the interaction time is much shorter. Nonetheless, \citet{Moore96,Moore98} showed that harrassment, the repeated high velocity encounters combined with the tidal field of the cluster, can affect cluster galaxies in the desired way. Finally, galaxy groups and clusters do not only contain galaxies, but also the intragroup  or intracluster medium (ICM). Especially cluster galaxies pass through this medium at high velocity, typically slightly supersonic (\citealt{Faltenbacher06}), and thus experience a substantial ram pressure (RP) on their gas disks (or gas halos). By comparing the average gravitational restoring force working on  the gas  disk to the  RP expected in massive clusters like Coma,  \nocite{Gunn72} Gunn \&{} Gott (1972, hereafter GG72) estimated that typical spirals should have their gas pushed out by the ICM head wind. Thus, the idea of ram pressure stripping (RPS) was born. 

Naturally, RPS is relevant not only for disk galaxies in clusters, but in a variety of contexts, e.g.~for elliptical galaxies (e.g.~\citealt{Lucero05,McCarthy08}) and dwarf galaxies in clusters and groups, and even dwarf galaxies in the gaseous halos of giant galaxies (e.g.~the Large Magellanic Cloud, \citealt{Mastropietro08,Mastropietro09,Haghi09}).

By comparing the different processes, \citet{Boselli06review} concluded that RPS is the best candidate to explain the observed differences between field and cluster spirals in the nearby universe. As shown below, RPS is able to remove gas from a galaxy in the desired way, the SFR and colour change in the aftermath. Besides being able to explain the trends in global properties, several galaxies are known to show detailed characteristics expected for RPS candidates (Sect.~\ref{sec:indiv}). 

RPS of disk galaxies in clusters was subject to intensive studies during the last decade. 
In this review, I will focus on recent work dedicated to RPS of massive disk galaxies in galaxy clusters. The emphasis is put on theoretical work, though I will frequently compare the theoretical results to observations.  I will start by summarising 
the results of basic theoretical work in Sec.~\ref{sec:basics}. Subsequent sections deal with different extentions to the basics in the effort to arrive at a coherent picture. Sec.~\ref{sec:summary} will summarise the current status.

\section{Basic theory: the gas removal process} \label{sec:basics}

\subsection{The Gunn\&Gott criterion for the stripping radius} \label{sec:gg}

In the simplest picture, the amount of gas lost from a ga\-la\-xy's disk is determined by the competing forces of RP and gravitational restoring force.  The case where the disk galaxy moves face-on through the ICM is easiest to access: Consider a disk galaxy in cylindrical coordinates, $(R,Z)$. In order to determine the restoring force for a given disk radius, $R$, one needs to find the gravitational acceleration in $Z$-direction, $a_Z$, i.e.~perpendicular to the disk. For a symmetrical potential, $a_Z$ directly in the disk plane is zero. Thus, for a given $R$, one can move up from the disk in $Z$-direction and find the maximum of $a_Z$ for this $R$. Multiplying this maximum gravitational restoring  acceleration, $a_Z{}\Max (R)$, with the gas disk's surface density, $\Sigma(R)$, yields the graviational restoring force per unit area, 
\begin{equation}
f\Grav(R) =  a_Z{}\Max (R) \cdot \; \Sigma(R) \label{eq:fgrav}
\end{equation}
at this disk radius $R$, which can now be compared to the RP. The restoring force per unit area is  a decreasing function of disk radius. For a given RP, the radius at which the restoring force equals the RP is called stripping radius. Inside this radius, the galaxy's gravitation outweighs the RP and thus can prevent gas removal, outside this radius the RP wins and strips the gas away. This method for estimating the stripping radius is commonly called Gunn\&Gott criterion (G\&G criterion, for short), although it contains two important improvements compared to the original version of GG72: First, it is spatially resolved instead of bein averaged over the whole galaxy. Second, the original calculation of the gravitational restoring force used the mass of the stellar disk only, whereas the method described above includes gravitational forces due to all components of the galaxy, i.e.~disk, bulge and dark matter halo. Like the original version, it is restricted to galaxies moving face-on through the ICM. 

The RP, 
\begin{equation}
p\Ram = \rho\ICM v\Gal^2,  \label{eq:pram}
\end{equation}
is determined by the density of the cluster gas, $\rho\ICM$, and the galaxy's velocity through this medium, $v\Gal$. Thus, RPS is expected to be most effective near cluster centres, where both,  $\rho\ICM$ and $v\Gal$, are highest. 

\subsection{Early simulations}
The first RPS simulations were dedicated to the removal of gas replenished by stars from elliptical galaxies (\citealt{Gisler76,Lea76,Shaviv82,Takeda84,Gaetz87, Portnoy93,Balsara94}).

The first simulations of disk galaxies (\citealt{Farouki80,Toyama80}) investigated whether RPS can convert gas rich spirals into gas poor S0. Despite low resolution and the restriction to two dimensions, already these simulations agreed that for typical cluster conditions RPS can strip typical spirals severely or completely, and that stripping works at the outer parts of the disk first.

\subsection{Systematic studies}
Later work (\citealt{Abadi99,Schulz01,Marcolini03,Roediger05,Roediger06}) performed more systematic studies at higher resolution, both, in 2D and 3D. Simulations were done mostly for massive disk galaxies, but also for dwarfs (\citealt{Marcolini03}), and mostly for ICM conditions representative for cluster centres, but also for low RP environments (\citealt{Roediger05,Roediger06,Marcolini03}).  The key question was to test the  G\&G criterion for the stripping radius. Generally, for face-on geometries, this simple criterion does a good job. 

Studying galaxies inclined w.r.t. the orbit requires 3D simulations.  Test cases were included by \citet{Abadi99,Quilis00,Schulz01}. \citet{Marcolini03} performed a systematic study for dwarfs, \citet{Roediger06}for giants. The common result is that inclination does not have a significant influence unless the galaxy moves close to edge-on. Mass loss in  face-on stripping (defined as inclination $0\degree$) up to an inclination of  $60\degree$  is very similar. The inclination is relevant mostly for medium RPs, that strip a galaxy severely, but leave some gas in the inner part of the disk. Strong RPs that strip a face-on galaxy completely, also strip an edge-on moving galaxy, although on a somewhat longer timescale. Weak RPs that hardly affect a face-on galaxy, neither strip an edge-on galaxy. 
The most prominent influence of inclination is that stripping of inclined galaxies makes the gas disk asymmetric (\citealt{Roediger06}).

\subsection{Timescales: ram pressure stripping as a multistage process}
More recent simulations (\citealt{Schulz01,Marcolini03,Roediger05,Roediger06})  which were run for longer than 100 Myr, revealed that RPS is actually a multi-stage process. The first stage could be called RP pushing: here the RP pushes the disk gas from disk radii larger than the stripping radius.  This displacement happens on short timescales of a few 10 Myr. 

However, the gravitational potential of a galaxy is deep and extends beyond the disk region. Consequently, accelerating the displaced disk gas to escape velocity requires an ongoing RP.  Unbinding the disk gas from the galactic potential takes a few 100 Myr.

In addition to the gas removal due to the RP, grid simulations can resolve the Kelvin-Helmholtz (KH) instability that is expected to work continuously on the gas disk's surface. Gas loss due to this continuous or turbulent-viscous stripping (\citealt{Nulsen82}) happens at a much lower rate ($\lesssim 1 M\Sun \Yr^{-1}$) and thus becomes evident in long-time simulations only.

\subsection{Common conventions and simplifications}
The "classic" RPS as proposed by GG72 happens only during the first of the phases described in the previous section, it is the same as the RP pushing. In recent years, also the complete multistage process is referred to as RPS. Alternatively, the complete process is called gas stripping or ICM-ISM interaction (ISM for interstellar medium). 

Simulations are mostly done in the galaxy rest frame, where the motion of the galaxy through the cluster translates into an ICM wind flowing past the galaxy. Therefore, the equation for the RP often reads $\rho\ICM v\ICM^2$, where the ICM wind velocity, $v\ICM$, replaces the galaxy's orbital velocity. 

Different methods are used to model the ICM-ISM interaction. The simplest are particle codes, where the RP is included as an additional force on gas particles. Hydrodynamical simulations utilise either smoothed particle hydrodynamics (SPH) or grid codes. Each method has its strengths and problems (e.g.~\citealt{Agertz07,Tasker08}).

Although the RP varies along the orbit through a cluster, the first step was to use constant ICM wind.  This simplification was used by all work discussed in  this Sec.~\ref{sec:basics}. The second important simplification is that of a homogeneous gas disk, although the interstellar medium (ISM) of real disk galaxies is highly inhomogeneous. On the observational side, the first indication regarding the gas contents was the observation that many cluster spirals have truncated gas disks. The homogeneous gas disks in the simulations are commonly meant to be comparable to the diffuse HI disks. 

The next sections describe different extentions to this basic work.

\section{Varying ram pressure during cluster passage} \label{sec:orbit}

When a galaxy passes through a galaxy cluster, both, its orbital speed and the ICM density surrounding it, increase towards the cluster centre and decrease after centre passage. Consequently, the temporal RP profile is peaked.

An obvious step to improve the basic simulations is to make the ICM wind vary accordingly. For elliptical galaxies, this was attempted since the 70ies (\citealt{Lea76,Takeda84,Acreman03}), although with restrictions to 2D, low resolution and radial orbits.  \citet{Toniazzo01} modelled the passage of an elliptical galaxy on a rosetta-like orbit through a Coma-like cluster in 2.5D. 

The constant ICM wind simulations for disk galaxies suggest interesting effects. Given that it takes a few 100 Myr to accelerate the gas displaced in the RP pushing phase enough to become unbound from the galaxy's potential, a short enough RP peak may be unable to unbind all gas, but a fraction of the displaced gas could fall back to the disk. This would mean a deviation from the G\&G criterion.

Such a fallback indeed takes place in the simulations of 
\citet{Vollmer01}, who were the first to model RPS of a disk galaxy in a varying RP. In this work, the gas disk was modelled by a sticky-particle code. The RP was included as an additional acceleration on particles at the upstream side of the galaxy. This description cannnot capture hydrodynamic effects like underpressure in the wake, eddies, or KH instabilities. The orbits through the cluster were approximated to be purely radial, i.e.~the RP does not vary in direction, 
allowing to use of an analytical temporal RP profile. The model cluster is tailored to resemble the Virgo cluster. As Virgo is very compact, the RP peak is rather short. In this case,  stripping becomes a distinct event, and indeed displaced gas falls back to the disk when the RP ceases. The authors propose that this feature is useable to distinguish between ongoing and recent stripping. 

\nocite{Roediger07} Roediger \&{} Br\"uggen (2007, hereafter RB07) and \citet{Jachym07} were the first to study the cluster passage with full 3D hydrodynamic simulations.  The SPH simulations of \citet{Jachym07,Jachym09} concentrated on galaxies orbiting on purely radial orbits through rather compact, Virgo-like clusters and thus also found RPS to be a distinct event, leading to backfall of displaced gas. 

\begin{figure}
\includegraphics[width=0.4\textwidth]{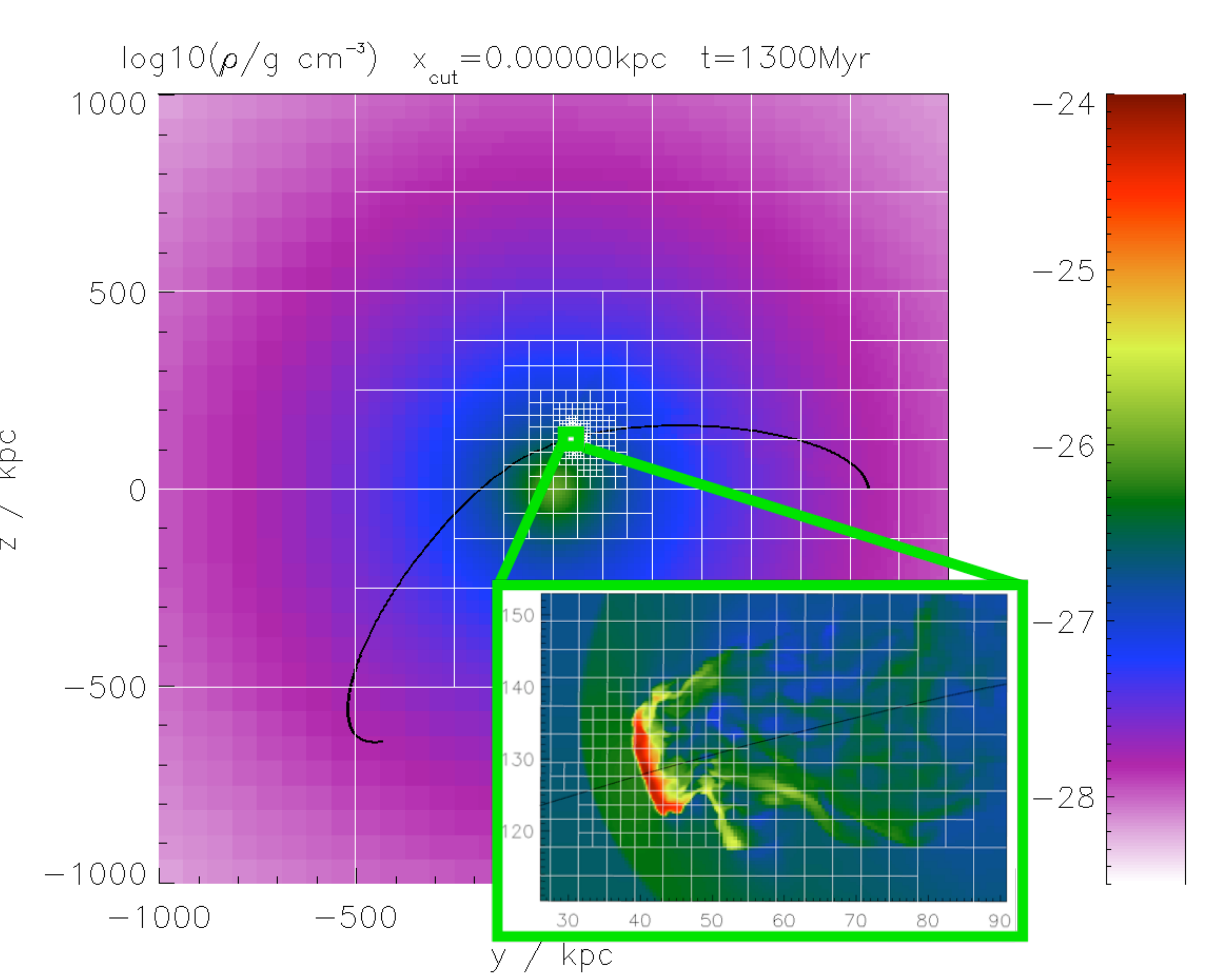}
\caption{RPS simulation of RB07,RB08.  Slice through the computational grid in the orbital plane, showing the colour-coded gas density. The background picture shows the whole cluster, the inset zooms on the galaxy.White boxes show blocks of $8^3$ grid cells. The galaxy was resolved with 250 pc.}
\label{fig:mysim}
\end{figure}

The grid simulations of RB07 were dedicated to a compact cluster (comparable to Virgo, though not as compact) and an extended, Coma-like cluster. 
These simulations were conducted in the cluster rest frame. While most of the cluster is covered in a low resolution grid, the adaptive grid always refines highly on the moving galaxy (see Fig.~\ref{fig:mysim}).
%
%
In the galaxy's wake, the resolution was kept on the 1-2 kpc level, allowing a study of RPS tails (see Sec.~\ref{sec:tails}). Only at distances to the galaxy larger than 100 kpc, derefinement was enforced to limit computational costs. Despite the lower resolution at larger distances from the galaxy, this method even allows to track the distribution of the stripped gas through the cluster. 
  In this more general case studied in these simulations, no backfall of gas specifically after core passage is seen. Only in case of a very rapid increase of RP, the gas loss is somewhat delayed compared to the expectation of a time-dependent G\&G criterion. One reason may be that the cluster studied by Vollmer et al. is more compact than the ones studied by RB07, another may well be the differing numerical methods.  
  
\begin{figure}
\includegraphics[angle=0,width=0.37\textwidth]{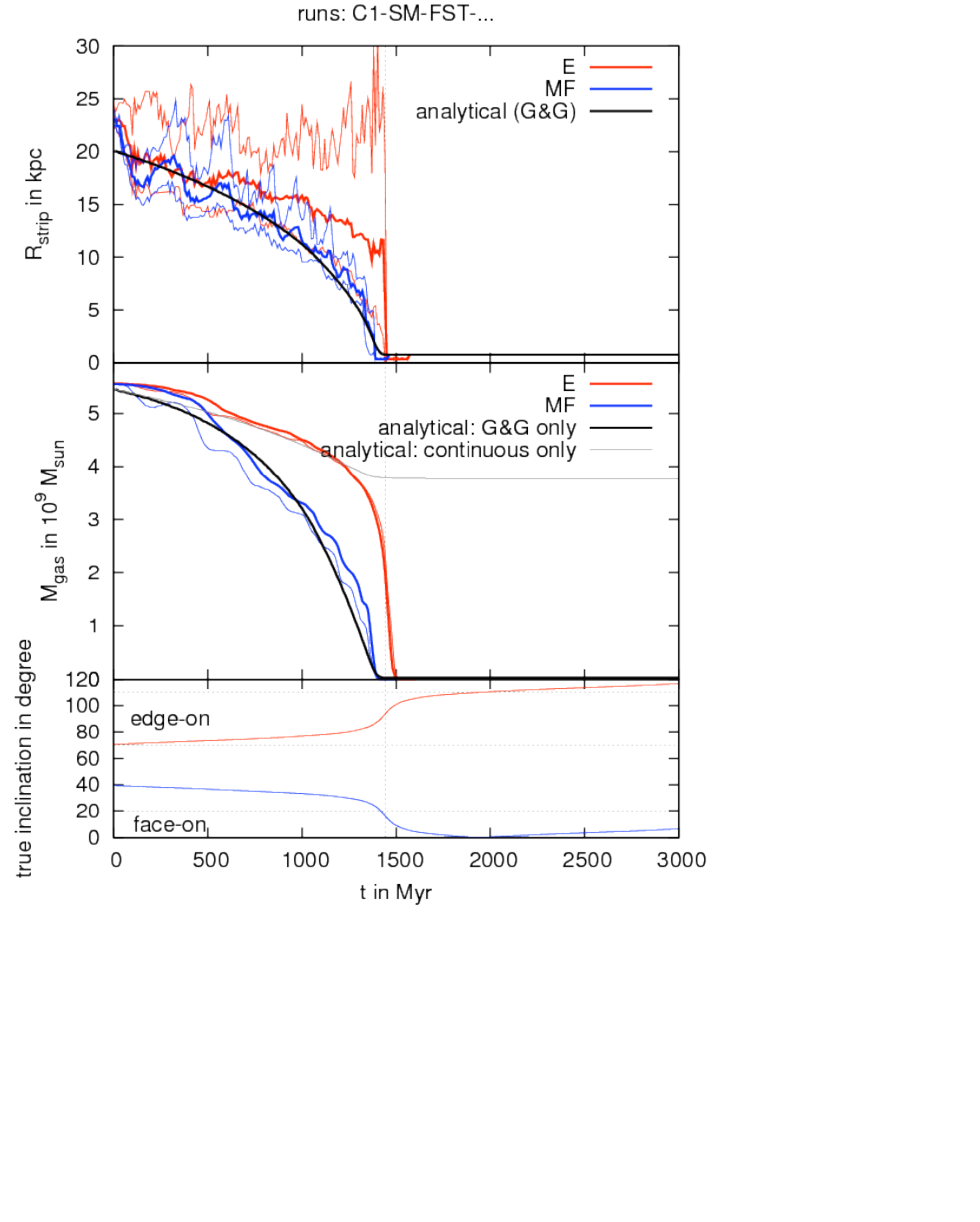}
\caption{Evolution of the gas disk radius (top, thick lines) and the gas mass (middle) during a cluster passage. The bottom panel displays the evolution 
of the true inclination, that is, the angle between the galaxyÕs rotation axis and the direction of motion. The red and blue lines are for two galaxies with different inclination. The thick black line is the estimate from the G\&G criterion. For the galaxy moving at medium to face-on inclination, the G\&G criterion describes the mass loss history well, for the edge-on galaxy it does not. The thin lines in the top panel show the maximum and minimum radius of the gas disk at each time. The edge-on galaxy is asymmetric, the other is not. From RB07.}
\label{fig:massplot}
\end{figure}

 A comparison of the simulated mass loss history with a time-dependent version of the G\&G criterion  revealed that this simple criterion is a reasonable approximation in the obvious situation: while the galaxy falls into the cluster, the galaxy should not move near edge-on, and the RP should not increase too rapidly. Otherwise simulations and analytical estimate differ (Fig.~\ref{fig:massplot}).
%
%
 Moreover, the mass loss due to KH stripping in long low RP or edge-on periods is not captured by the G\&G criterion. 

Considering RP histories for various orbits in the two clusters reveals interesting implications for the cluster ga\-laxy population (RB07): The constant ICM wind simulations distinguished strong ($\gtrsim 10^{-10}\Presunit$), medium ($\sim 10^{-11}$ $\Presunit$), and weak  ($\lesssim 10^{-12}\Presunit$)  RP re\-gimes, where a typical massive spiral (rotation velocity 200 $\Kms$) would be stripped completely, seriously (say down to a few kpc gas disk radius) or just marginally. In realistic clusters, no orbits with only weak RPs are possible. Even in the compact cluster on orbits with large impact parameters, a galaxy experiences at least a medium RP. In order to move on a more circular orbit, a galaxy needs to have a higher orbital veloctity, which results in a medium RP despite a low ICM density. For radial orbits, the high central ICM density together with the high central velocity nearly always leads to complete stripping. Especially in the extended, Coma-like cluster, typical spirals are stripped completely before even reaching the cluster centre. \citet{Bruggen08} investigated RP histories of cluster galaxies in more detail by analysing the Millenium Simulation (\citealt{Springel05}), finding that about one quarter of ga\-laxies in massive clusters are currently subject to strong RP, and more than 64\% of current cluster members have experienced strong RPs since they fell into their host cluster.

\section{Studies on individual galaxies} \label{sec:indiv}

\subsection{Observable signatures of ram pressure stripping}
The main impact of RPS on a disk galaxy is the truncation of its gas disk down to the stripping radius, whereas the stellar and dark matter components are left completely unaffected. In contrast, gravitational interactions tend to influence all components of the involved galaxies. Thus, an undisturbed stellar disk combined with a truncated gas disk is commonly interpreted as a sign that this galaxy has suffered RPS.  Additionally, the stripped gas should form a tail on the downstream side of the galaxy and thus make it possible to determine the galaxy's direction of motion.  Rotation curves and 2D velocity fields of the remaining gas disks are expected to be only weakly effected by RPS, and only in near edge-on cases (\citealt{Kronberger08}).

\subsection{Examples}
Especially Vollmer et al. (1999\nocite{Vollmer99} - now, for a summary see \citealt{Vollmer09holistic}) have done extensive work in observing candidate galaxies deeply in HI, radio and other wavelengths. Additionally, they modelled each candiate with a sticky particle code (\citealt{Vollmer01}), aiming at disentangling each galaxy's PRS history and stage. The authors stress that this aim cannot be reached by comparing projected gas density maps alone, but velocity information is essential. Although this method is applied successfully to a number of galaxies, one needs to keep in mind that the applied numerical method lacks some essential physics. E.g.~the fallback of not fully stripped gas after core passage is used as a criterion to distinguish ongoing and recent RPS, the use of this criterion is not supported by hydrodynamical simulations (RB07). 

 The (incomplete) list of examples given here contains members of Virgo, the most nearby and hence best-studied cluster. This list focusses on the remaining (gas) disks, RPS tails are discussed in Sec.~\ref{sec:tails}.

\begin{figure*}
\includegraphics[height=4.7cm]{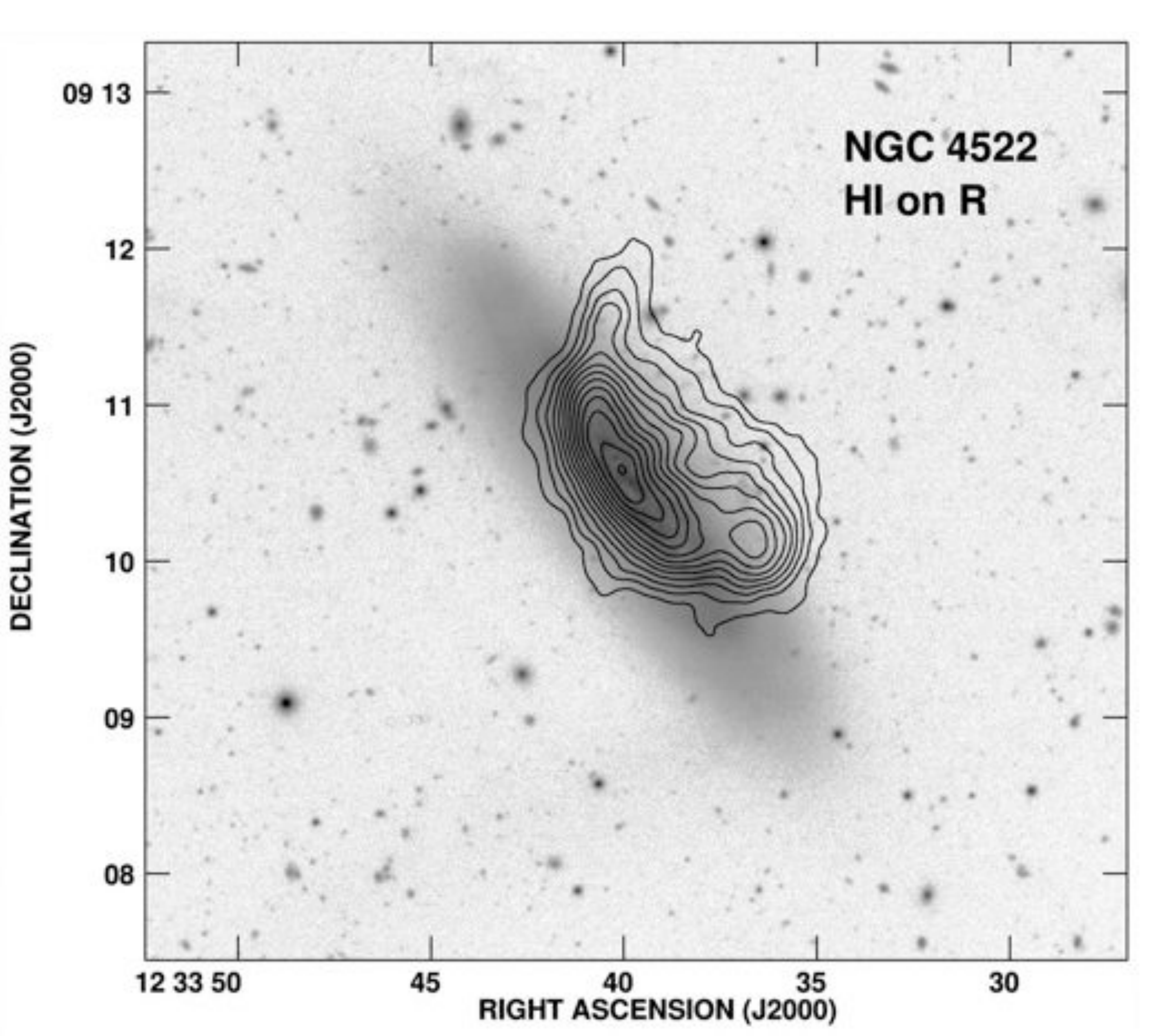}
\includegraphics[height=4.7cm]{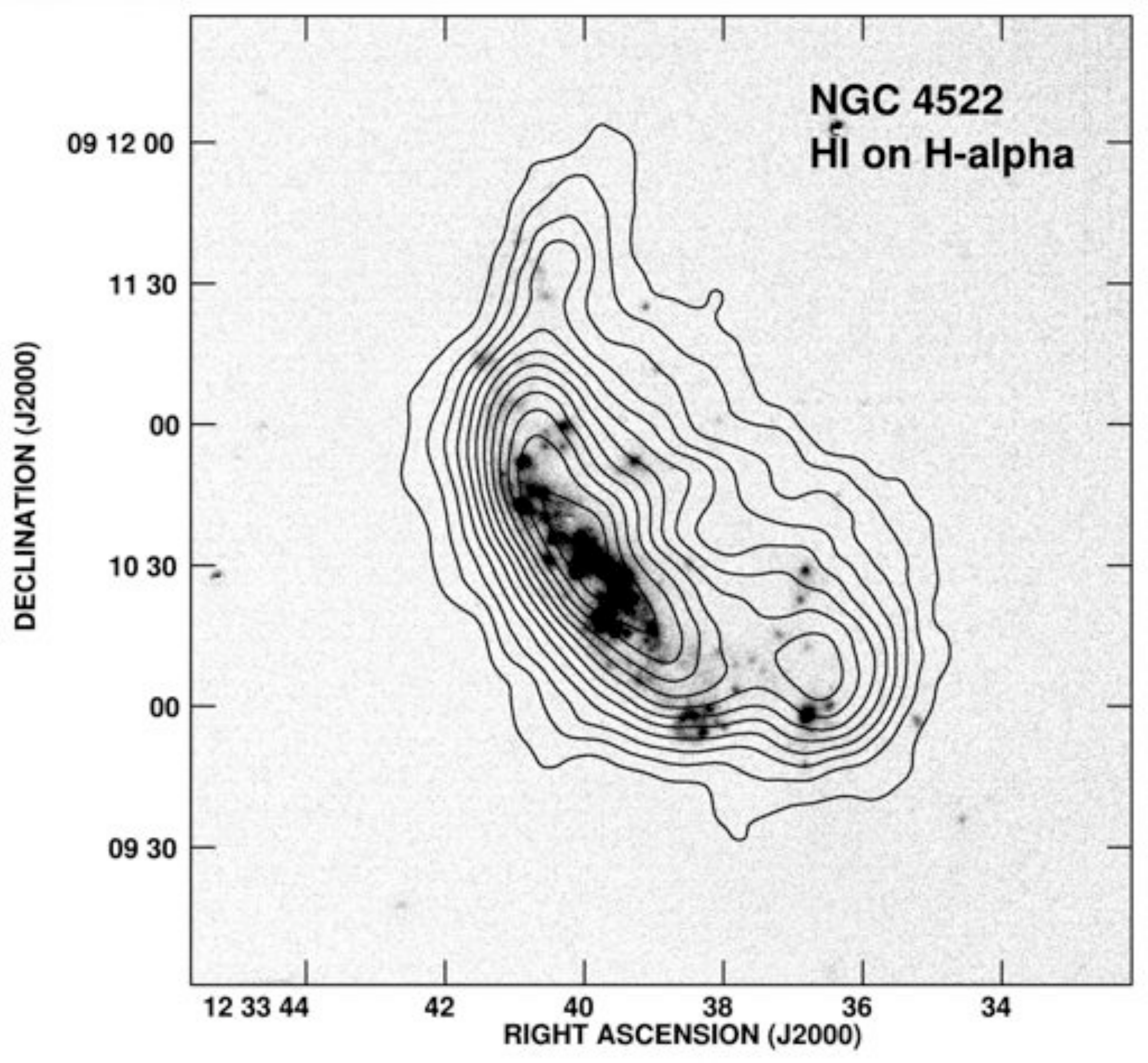}
\includegraphics[height=4.9cm]{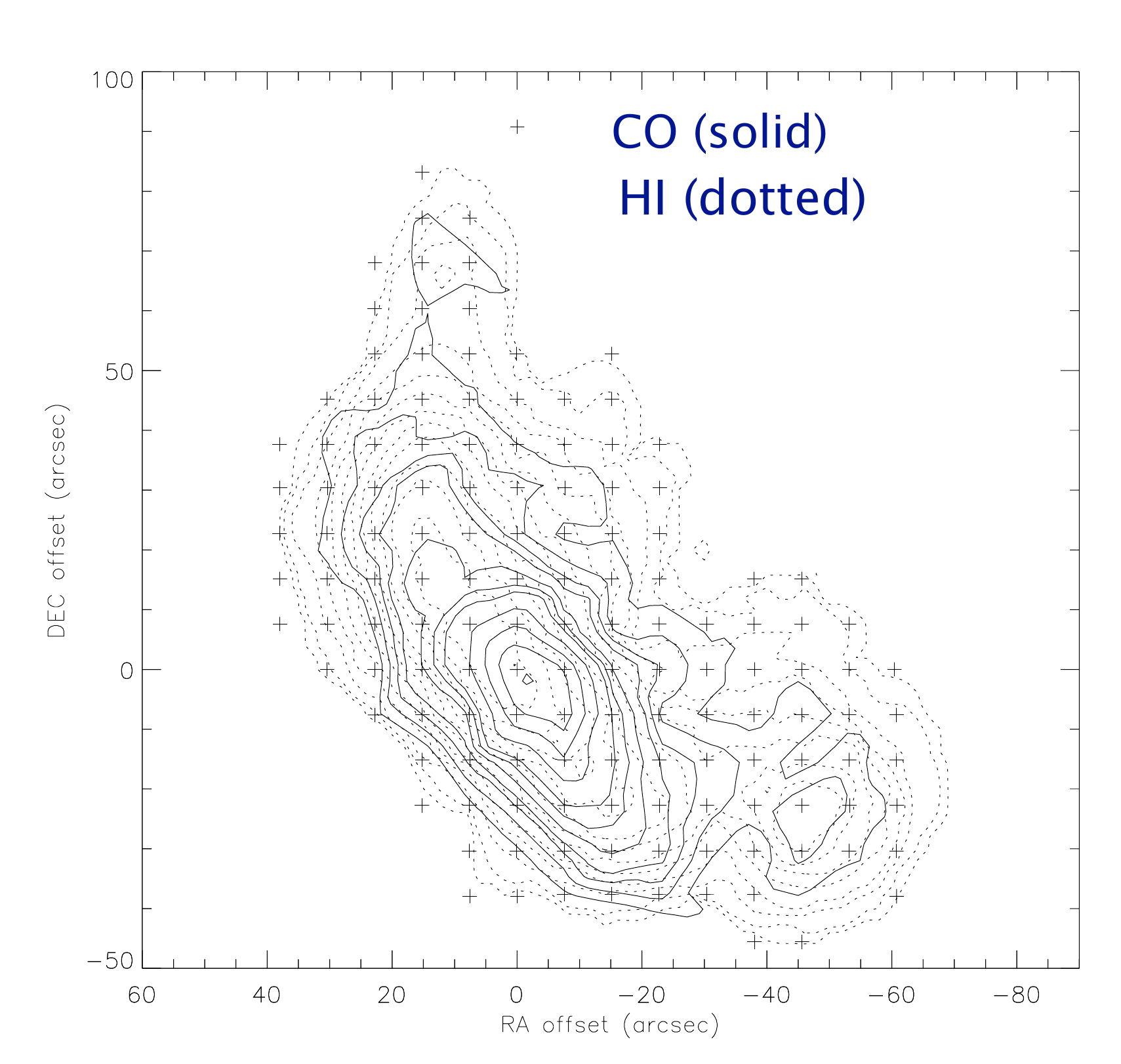}
\caption{
Textbook example NGC4522: an undisturbed stellar disk, but truncated HI disk and extraplanar gas. The H$\alpha$  and CO disk are truncated consistently with HI. 
Stellar disk: R-band, greyscale in left panel. HI: contours in left and middle, dotted right panel. H$\alpha$: greyscale middle panel. CO: solid contours in right panel. 
 (Left and middle from \citealt{Kenney04}, right from \citealt{Vollmer08}). 
 }
\label{fig:ngc4522}
\end{figure*}

\paragraph{NGC4522} is one of the best-studied RPS candidates, data in many wavebands is available.  All wavebands detecting ISM (HI:  \citealt{Kenney04}), H$\alpha$: \citealt{Kenney99,Vollmer00},  CO: \citealt{Vollmer08},  (polarised) radio continuum:\citealt{Vollmer04ngc4522}, far infrared: \citealt{Murphy09,Wong09}) consistently show  a gas disk truncated at 3 kpc disk radius, whereas the stellar disk is mainly undisturbed (Fig.~\ref{fig:ngc4522}).
%
%
Moreover, significant fractions of extraplanar H$\alpha$ and HI are found, suggesting a tail of  3kpc length. The polarised radio continuum suggests compression of the ISM at the upstream edge. From optical and UV data, \citet{Crowl06} show that the SF just outside the stripping radius ceased  just $\sim 100$ Myr ago, after a modest star burst.

Comparing sticky particle+MF advection (see Sec.~\ref{sec:mf}) simulations to these observations, \citet{Vollmer06} conclude that this galaxy currently suffers RPS, or has experienced the RP maximum just 50 Myr ago. 

Thus, NGC4522 appears to be a textbook example of RPS. However, the needed RP to produce the observed stripping is unusually high given the large cluster-centric distance  (800 kpc) of this galaxy. \citet{Kenney04} and \citet{Vollmer06} suggest that local enhancements in ICM density or bulk flows may be responsible for the enhanced RP at this galaxy's position, a plausible explanation, because the Virgo cluster is still in the stage of assembly. Investigations of cosmological simulations (\citealt{Tonnesen08,Ludlow09}) support this idea by revealing that at a given clustercentric distance, galaxies can experience RPs of a range up to or more than one order of magnitude, and that cluster galaxies can have unorthodox orbits. 

\paragraph{NGC4402} shows a truncated HI and radio continuum disk. Moreover,  ablation of dense clouds seems to be at work (see \citealt{Crowl05} and references therein).

\paragraph{NGC4430} shows truncated HI (\citealt{Chung07}), H$\alpha$, near UV, IR and radio continuum, and radio deficit at the upstream edge (\citealt{Abramson09}). From simulations, \citet{Vollmer09holistic} suggests pre-peak stripping. No significant enhancement of SF is found, and only very little SF in the tail.

\paragraph{NGC4569} was studied extensively by \citet{Boselli06} and others (e.g.~\citealt{Wilson09,Chyzy06, Vollmer04}). Also for this galaxy data at many wavelengths is available. From multi-wavelength data, \citet{Boselli06} date the removal of the outer gas disk to to have happened 100 to a few 100 Myr, in agreement with the dynamical model of \citet{Vollmer04}.

\section{Ram pressure tails} \label{sec:tails}

Whereas the physical picture for the gas removal appears to have converged, the fate of the stripped gas is less clear, both from theoretical and observational point of view. 

\subsection{Theoretical work}

First theoretical attempts were again made for elliptical ga\-la\-xies (\citealt{Stevens99,Toniazzo01,Acreman03}), predicting weak, but observable X-ray wakes. The basic studies on disk galaxies naturally see the stripped gas forming a tail on the galaxy's downstream side. 

Studies aiming at the morphology of the tails, however, need to consider a number of issues generally not included in the work discussed so far: (i) Studying the tail requires a larger volume in full 3D to be simulated with sufficient resolution, which causes higher computational expenses for both, SPH and grid simulations, and thus was not done in most cases. (ii) The tail structure will depend on the RP history.   (iii) The tail structure will depend on additional physical properties of ICM and ISM, like viscosity, thermal conduction, cooling, self-gravity, SF, and MFs. Approaches so far have realised only subsets of these requirements. 

\begin{figure}
\includegraphics[width=0.24\textwidth]{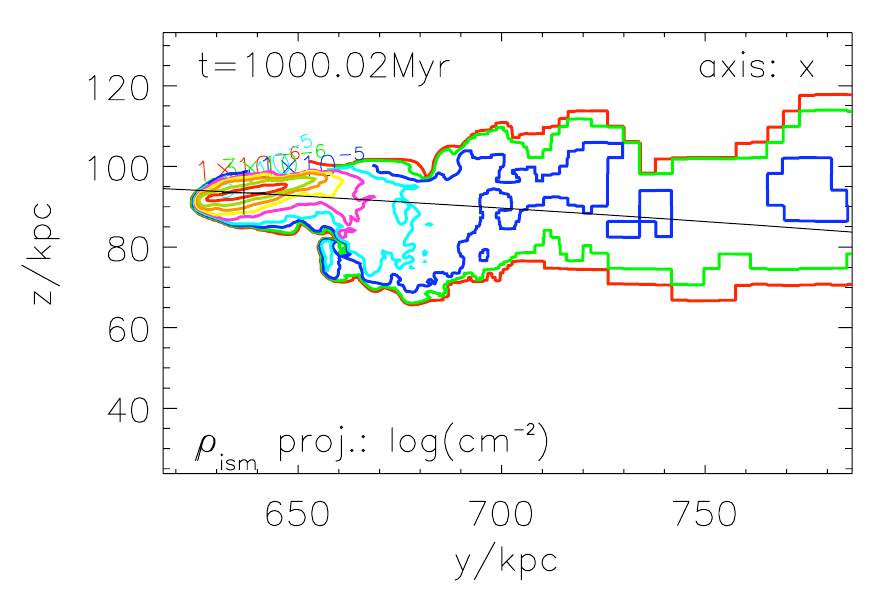}
\includegraphics[width=0.24\textwidth]{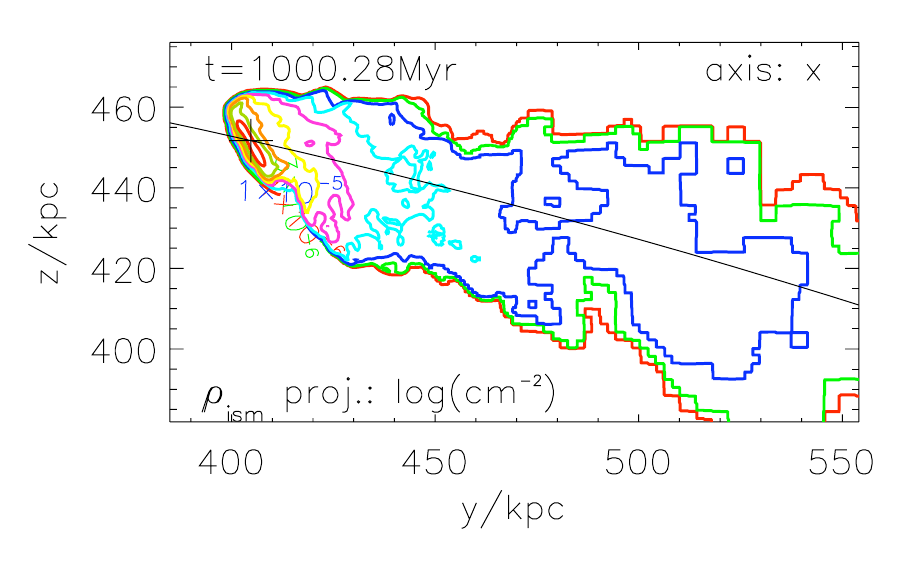}
\caption{Simulated RPS tails (RB08):  projected gas density contours.  Contour spacing is logarithmic, half an order of magnitude per step. The cyan coloured (the 4th from outside) contour  corresponds to $\sim 2 \cdot 10^{19} \CM^2 $. The black line marks the orbit.}
\label{fig:mysim_tails}
\end{figure}

The first study dedicated to RP wakes (\citealt{Roediger06wakes}) found flaring tails of stripped gas, which must however be interpreted very carefully, as a constant ICM wind was used. The subsequent simulations (RB07, \citealt{Roediger08}, hereafter RB08) modelled the passage of a galaxy though a cluster in a 3D adaptive grid code (Fig.~\ref{fig:mysim}).  
Assuming that the simulated gas disk is HI gas and remains so after stripping, mock HI maps can be produced (Fig.~\ref{fig:mysim_tails}).  
%
%
For a sensitivity limit of $~\sim 10^{19}\CM^{-2}$ in projected gas density, typical tail lengths are  $40\Kpc$. Such long tails are seen even at
large distances (0.5 to $1\Mpc$) from the cluster centre.  
Tail length and density are not necessarily largest  in the cluster centre, where the RPs is strongest. Higher RP causes a higher gas loss per time. However, higher RPs are also accompanied by high velocities. Consequently, the stripped gas is distributed over a larger volume. The mass loss per orbital length, which depends on RP and orbital velocity, determines density and (observable) length of the tail. 

Using an SPH code including gas cooling and recipes for SF, but a constant ICM wind, \citet{Kapferer09} also produce mock observations.
Their simulated tails tend to be very long, up to a few 100 kpc. Part of the gas in the tail continues to cool and form dense knots and stars all along the tail. 

Although the current simulations seem to be able to reproduce aspects of the observations (see below), they should be interpreted cautiously. E.g.~the physics of SF itself is not yet fully understood, and the relevance of  MFs and transport processes remains to be investigated.

\subsection{Comparison to Observations}

Also the observational situation remains unclear. Given the many galaxies with truncated HI disks, at least some HI in tails or simply distributed through the clusters could be expected. However, this gas is not found easily. \citet{Vollmer07where} searched the vicinity of a number of known RPS candidates deeply and found no more HI than already known. A blind HI survey (\citealt{Kent07,Gavazzi08}) found only a handful HI clouds unassociated with galaxies. One  of them could be interpreted as RP stripped gas (\citealt{Kent09}). A deep targeted survey of Virgo spirals (\citealt{Chung07}) found HI tails for galaxies at distances between 0.6 and 1 Mpc from the cluster centre. Comparable cases were found in the simulations of RB08, which showed a similar structure in projected gas density, but generally the simulated wakes show a much more turbulent wake than observed. 
At first glance, long tails at such large cluster-centric distances are surprising, because RPS is strongest near the clusters centre. However, as discussed above, the observability of the tail depends on the gas loss per orbital length, not RP alone.  As also discussed above, galaxies at a given cluster-centric radius can experience a whole range of RPs, so these examples could be fast-moving galaxies.

Near NGC4388, a Virgo spiral,  \citealt{Oosterloo05} observe a 120 kpc long HI cloud that appears to be, at first glance, a RPS tail. However, neither the tail structure in the HI image nor the velocity structure along the tail can be explained by the current simulations (RB08). Whether this means that this tail is not RP induced or the simulations are insufficient, remains to be seen.

In addition to HI tails, there are
only very few galaxies that have X-ray (\citealt{wang04,Sun06}), 
H$\alpha$  (\citealt{gavazzi01,Sun07,Yagi07}) or radio tails (e.g.~\citealt{Gavazzi95}).  
These tails tend to be rather long
(several 10 kpc), and extremely narrow ($< 10\Kpc$) and straight. Moreover, especially the
X-ray tails seem to be rare: \citet{Sun07a} have searched for additional cases
in the Chandra and XMM data of 62 galaxy clusters and did not find any. 
The few known X-ray
and H$\alpha$ tails are generally much narrower and much straighter than the simulated
tails. Thus, additional physics like a viscous ICM,
the influence of cooling and tidal effects may be
needed to explain the observations.

\section{Ram pressure stripping of a multiphase interstellar medium  and the effect on star formation} \label{sec:multiphase}

It is commonly argued that RP stripped galaxies will also cease to form stars and thus subsequently become dead and red, because the galaxy has lost its gas reservoir, which would be needed for further SF activity. However, the simulations discussed so far mostly dealt with the diffuse gas phases of the ISM (diffuse HI, ionised) only. A realistic ISM is highly inhomogeneous and contains also HI and molecular clouds. As stars form in molecular clouds, the link from loss of diffuse gas to SF quenching has to be made very carefully via a multiphase ISM. 

Modelling RPS of a multiphase gas disk poses quite a challenge.  A variety of physical processes needs to be included: heating and cooling of the ISM, self-gravity, SF and stellar feedback, MFs.  An enormous range of spatial scales needs to be resolved, starting from subparsecs to resolve molecular clouds, the sites for SF, to at least 100 kpc, the size of RP tails. Given that the process of SF itself is still a matter of ongoing research (e.g.~\citealt{Bate09,Price09,Krumholz09}), and given limits for computational expenses, current simulations can only be approximations.

The question of RPS of a multiphase ISM has two sides: how does RPS differ between a homogeneous gas disk and an inhomogeneous one, and how is the multiphase ISM and thus the galaxy's SFR affected by RPS. 

\subsection{Gas loss from a multiphase disk}

The first-glance expectation for the first question is that an inhomogeneous disk will be stripped more easily, because the ICM wind blows away the low surface density parts of the disk, creating holes. This in turn enlarges the surface of the disk, which is more vulnerable to KH instabilities. RPS should be more efficient.  However, in the light of non-linear internal dynamics of ISM, one could argue towards an opposite answer: The increasing ICM pressure (static+RP) may compress ISM clouds, increasing surface density, making them harder to strip. Also MFs may prevent KH instabilities. A first attempt to find an answer by simulations was done by \citet{Quilis00}, who artificially cut holes into an otherwise smooth gas disk. No cooling was included, thus the disk is stripped more easily. A more advanced attempt was made by \citet{Tonnesen09}, who modeled an inhomogeneous ISM including self-gravity, cooling, and heating, with a resolution of  40 pc. Low surface density gas is stripped easily from any radius, although the overall mass loss does not differ much from models with a smooth disk. Interpreting the morphology of the remaining gas disk had to be done carefully, as runs with different resolutions lead to different results. 

Answering this question observationally is rather difficult, as for no galaxy the true velocity w.r.t. the ICM is known. 

\subsection{Effect of ram pressure stripping on star formation}
All observations and simulations agree that at outer disk radii, where the diffuse gas is stripped, also SF ceases. Although the stars form in  dense clouds, the internal dynamics of the ISM connect the cloudy and diffuse phase and thus SF to the diffuse phase. 

For the remaining disk, the answer is not so easy. The first-glance expectation is that the high ICM pressure leads to compression of the ISM and thus increased SF. Earlier work (\citealt{Fujita99,Bekki03}) indicates an increase in SFR by a factor of  a few. A similar or even stronger effect is seen by \citet{Kronberger08sf} and \citet{Kapferer08,Kapferer09}, who utilise an SPH code including cooling and standard recipes for SF and stellar feedback.  Interestingly, stars are also formed in the stripped gas.  However, their model galaxies experience a constant RP. Thus, the evolution of the SFR in a varying RP is still an open question. 

Observations do not give a clear answer, either. A study on H$\alpha$ properties of Virgo spirals (\citealt{Koopmann04,Koopmann04sf,Koopmann06,Koopmann06atlas}) suggests that the SF in the outer disks, where gas was stripped, ceases, whereas the SFR in the remaining inner disk is moderately enhanced. However, other observations do not find an increased SF despite clear RPS signatures (NGC4330 \citealt{Abramson09}).

\section{Magnetic fields} \label{sec:mf}

\subsection{Magnetic fields in spiral galaxies}
Evidence for MFs comes mainly from (polarised) radio continuum emission
(i.e.~syncrotron emission of cosmic ray electrons spiralling around MFs) and
Faraday rotation measurements. The total intensity of the radio continuum
reveals the total strength of the MF in the plane of the sky. The polarised intensity
and the polarisation angle reveal the strength and structure of regular fields
in the plane of the sky. The observation of MFs is still very difficult, but
huge improvements in the quality of the data is expected from the next
generation of radio telescopes, e.g.~LOFAR (LOw Frequency ARray, 
and SKA (Square Kilometer Array%
).

Typically, spiral galaxies have a regular and a tangled MF
component (\citealt{beckLNP,beck05}; \citealt{beck07} and references therein).
The typical total field strength is around $10\MicroG$. The strength of
resolved regular fields is a few $\MicroG$. The regular field follows
the galactic plane and the spiral arms, although it is slightly offset. There
are also examples of regular fields that follow the interarm regions.
Occasionally, MFs sticking out of the galactic plane are found, usually they
are associated with supernovae outbursts and galactic winds. The MF in (gas)
spiral arms is mostly tangled. It seems to be connected to the turbulence
induced by SF.  The origins of galactic MFs are not fully
understood. A frequently discussed explanation comes from dynamo models, where
a seed field is amplified (see e.g.~review by \citealt{kulsrudLNP} and
references therein). They can explain a toroidal-like regular field to some
degree. 

%
\subsection{Magnetic fields in clusters} \label{sec:mf_ICM}
In the ICM, evidence for MFs comes from Faraday rotation measurements, inverse
Compton scattering of cosmic microwave background photons, and radio halos.
Typical MF strengths are a few $\MicroG$
(\citealt{ensslin05,feretti04}, and references therein). There is some evidence
that fields are intermittent and turbulent with
autocorrelation lengths of a few kpc in cluster centres and
a few 10 kpc in cluster outskirts. Also here, the MFs are probably
closely linked to ICM turbulence.

Besides direct measurements, there is plenty of indirect evidence for the
existence of MFs in the ICM. E.g.~bubbles rising buoyantly in the ICM seem to
be more stable than predicted by purely hydrodynamical simulations. Viscosity
or MFs with coherence lengths larger than the bubble size could stabilise them
by preventing hydrodynamical instabilities (\citealt{reynolds05,Ruszkowski07bub}).
%
%
\subsection{Magnetic fields and ram pressure stripping} \label{sec:intro_mf_rps}
In interacting galaxies, MFs trace regions of gas compression, strong shear
motions, and enhanced turbulence (\citealt{beckLNP}).   Such motions can be
caused by tidal interactions as well as interaction with the ICM.

Typical features of RPS candidates are
asymmetrical distributions of the polarised radio intensity,
with polarisation maxima at the upwind side of the galaxy
(\citealt{Wezgowiec07,Vollmer07} and references therein). Thus, the MFs
contain information about the velocity components in the plane of the sky, which is
not accessible otherwise.  Moreover, distorted MF configurations have been
found in galaxies that show no other traces of interactions. Thus, MFs could
be either sensitive even to weak RPs that leave no other observable
traces, or they may remember distortions that happened several 100 Myr ago
(\citealt{Otmianowska03,Wezgowiec07}).

The question of MFs and RPS again has two sides:  The influence of RPS on MFs in spirals is addressed in a series of papers based on the method described in \citet{Otmianowska03}. First, RPS of a disk galaxy is modelled by the sticky particle method of \citet{Vollmer01}. In a manner of post-processing, the initial galaxy is given a toriodal MF, which is then evolved by solving the induction equation with an MHD grid code using the velocity fields produced by the sticky particle code. Thus, the MF is advected along with the
gas, but no effect of the MF on the dynamics of the gas is taken into
account. Despite problems in mapping between the particle and grid code, this method can explain succesfully the enhanced polarised radio contours found at the upstream edge of  RPS galaxies (\citealt{Wezgowiec07,Vollmer07}).  This procedure is applied to a number of RPS candidates (e.g.~NGC4522: \citealt{Vollmer06},  NGC4654: \citealt{Soida06,Soida06ngc4654},   NGC4254:  \citealt{Chyzy06,Chyzy08})

Simulations of the motion of a gas cloud through a magnetised ICM show the effect of draping (\citealt{lyutikov06,Jones96,Dursi08,Asai04apj,Asai05,Asai07,Ruszkowski07bub}): the MF is swept up and wrapped around the cloud, suggesting that MFs will also have an influence on the stripping of disk galaxies. 

A self-consistent approach is still missing.

\section{Intracluster medium properties} \label{sec:visc}

The ICM is an ionised medium, where protons and electrons have differing mean free paths. Strictly speaking, they should be treated as two different fluids. \citet{Portnoy93} investigated the differences between two-fluid and single fluid treatments for RPS of elliptical galaxies and found differences in the temperature structure of the wake.

Due to the strong temperature dependence of the viscosity and the high ICM temperature,  in the unmagnetised case, the ICM becomes highly viscous, so that flows on scales of galaxies should be viscous. 
Also based on observations of the Perseus cluster, it has been suggested by \citet{Fabian03,Fabian03a}, that viscosity may play an important role in dissipating
energy injected by the central AGN.  Circumstantial evidence for the presence
of significant ICM viscosity is also provided by an examination of the
morphology of H$\alpha$ filaments in the Perseus cluster. Several of the
filaments appear to trace well-defined arcs which argues against the presence
of strong turbulence in the ICM core, possibly resulting from the action of
viscosity.

On the other hand, even weak MFs
lead to a tiny proton gyroradius, which results in a very efficient
suppression of transport processes like viscosity and thermal conduction. If the fields are tangled (\citealt{clarke04,ensslin05}) on small enough scales, also a fluid description is justified, and the macroscopic viscosity may be greatly reduced.

In summary - the properties of the ICM are poorly constrained (\citealt{McNamara07}). In order to study the possible impact of a viscous ICM on RPS, \citet{Roediger08visc} simulated stripping of a disk galaxy in an ICM of various viscosities. Including viscosity requires a very small timestep, hence these simulations were done in 2D.  The stripping radius and amount of lost gas was not affected by the viscosity. However, as expected, the more viscous the ICM, the smoother the tail of stripped gas. The 2D restriction prohibits, however, a detailed interpretation of the tail structure.

\section{Enrichment of the intracluster medium by ram pressure stripping} \label{sec:enrich}

The gas lost from a RP stripped galaxy becomes part of the ICM, enriching the ICM with metals. One or more of such enrichment processes are expected to take place, because the ICM is not primordial but has metalicities of 1/3 to 1/2 solar. Besides RPS, also galactic winds, outflows from active galactic nuclei, and intracluster stars contribute to the enrichment. Simulations (see recent reviews by \citealt{Schindler08,Borgani08}) find that these enrichment processes lead to an inhomogeneous metal distribution. E.g.~RPS preferably pollutes cluster centres, whereas galactic outflows can be suppressed by the high ICM pressure in centres and are more effective in the outskirts. On top of the radial variation of the metallicity in clusters, both, observations and simulations (see review by \citealt{Werner08}), show a rich substructure of clumps and filaments. Apparently, the lost gas  is not mixed with the ICM immediately. According to the simulations, no proposed enrichment process can produce all observed metals, a combination of  processes is needed. The upcoming generation of X-ray observatories will be able to produce highly resolved metallicity maps, which will help to constrain and understand the various gas removal processes.

\section{Discussion and summary} \label{sec:summary}

Early work as well as more recent simulations agree that in many situation the gas removal from disk galaxies by RP can be approximated well by the simple analytical G\&G criterion (Sec.~\ref{sec:gg}). The estimate works best for galaxies moving face-on, and still gives reasonable results for inclinations deviating up to $60\degree$ from the face-on case. It fails for near-edge on symmetries. Even a time-dependent version of the G\&G criterion can be applied to galaxies experiencing a varying RP along their orbits through a cluster, unless the cluster is very compact (Sec.~\ref{sec:basics}). Hence, RPS seems to be able to explain the truncated gas disks found in many cluster spirals (Sec.~\ref{sec:indiv}).  When RPS removes the gas from the outer part of a galactic disk, also SF ceases in the stripped parts. This is a plausible idea, and finds general observational support.

Beyond these basics, there is a number of recently studied but still unsolved aspects:
\begin{itemize}
\item Several studies (observational and theoretical) suggest that RP causes an enhancement in SF activity in the remaining gas disk (Sects.~\ref{sec:indiv}, \ref{sec:multiphase}). However, contradicting observations  exist, and theoretical work needs confirmation due to the complex nature of SF physics. 
\item RPS should produce gas tails behind the stripped galaxies. Some examples are observed, but current simulations cannot explain the properties of the tail satisfactorily (Sec.~\ref{sec:tails}).
\item The difficulties with the RP tails could arise due to missing physics in the models. First extensions towards including MFs (Sec.~\ref{sec:mf}) and transport processes (Sec.~\ref{sec:visc}) are made, but need improvement. Further issues include the temperature and ionisation structure inside the tails.
\end{itemize}

Advances in these open questions will also assist other astrophysical topics: 
The understanding of the tails of RP stripped galaxies will require the inclusion of additional phy\-sics, most likely ICM transport processes and MFs. In turn, success in this question will also help to understand and constrain ICM properties, which are relevant for a variety of other phenomena, e.g. the interaction between AGN jets and the ICM. The question of enhanced SF activity due to RP may be used as a test case to check our understanding of SF in a special situation.

\acknowledgements
ER thanks Marcus Br\"uggen for stimulating discussions and helpful comments, and  acknowleges the support of the DFG under grant BR2026/3 within the DFG Priority Programme 1177 `Galaxy Evolution' and the supercomputing grants NIC 2195 and 2256 by the John-Neumann Institut J\"ulich. 


\bibliographystyle{aa_noNote}
\bibliography{newbib_clean,hydro_processes,icm_conditions,galaxies}

\end{document}